\documentclass[structabstract]{aa}  
\usepackage{graphicx}
\usepackage{natbib}
\usepackage{color}
\def\mearth{{\rm M_\oplus}}
\definecolor{viol}{rgb}{0.5,0,1}

\newcommand{\modifn}[1]{\textcolor{black}{#1}}

\def\modif{\modifn}

\begin{document}
   \title{Composition and fate of short-period super-Earths}

   \subtitle{The case of CoRoT-7b}

   \author{D. Valencia\inst{1}
       \and
       M. Ikoma\inst{1,2}
       \and
     T. Guillot\inst{1}
     \and
   N. Nettelmann\inst{3}}

   \institute{Observatoire de la C\^ote d'Azur, Universit\'e de Nice-Sophia Antipolis, CNRS UMR 6202, BP 4229, F-06304 Nice Cedex 4, France
     \and
     Dept. of Earth and Planetary Sciences, Tokyo Institute of
     Technology, Ookayama, Meguro-ku, Tokyo 152-8551, Japan
     \and
Institut f{\"u}r Physik, Universit{\"at} Rostock, D-18051 Rostock, Germany}


 \abstract
{The discovery of CoRoT-7b, a planet of radius $1.68\pm0.09\,\rm R_\oplus$, \modif{mass $4.8\pm0.8, \rm M_\oplus$} and orbital period of $0.854$\, days demonstrates that small planets can orbit extremely close to their star.}
{Several questions arise concerning this planet, in particular concerning its possible composition, and fate.}
{We use knowledge of hot Jupiters, mass loss estimates and models for the interior structure and evolution of planets to understand its composition, structure and evolution.}
{\modif{The inferred mass and radius of CoRoT-7b are consistent with a rocky planet that would be significantly depleted in iron relative to the Earth. However, a one sigma increase in mass ($5.6\,\mearth$) and one sigma decrease in size ($1.59\,\rm R_\oplus$) would make the planet compatible with an Earth-like composition (33\% iron, 67\% silicates). Alternatively, it is possible that CoRoT-7b contains a significant amount of volatiles. For a planet made of an Earth-like interior and an outer volatile-rich vapor envelope, an equally good fit to the measured mass and radius is found for a mass of the vapor envelope equal to 3\% (and up to 10\% at most) of the planetary mass. Because of its intense irradiation and small size, we determine that the planet cannot possess an envelope of hydrogen and helium of more than 1/10,000 of its total mass.}
We show that a relatively significant mass loss $\sim 10^{11}\rm\,g\,s^{-1}$ is to be expected and that it should prevail independently of the planet's composition. \modif{This is because the hydrodynamical escape rate is independent of the mean molecular mass of the atmosphere, and because given the intense irradiation, even a bare rocky planet would be expected to possess an equilibrium vapor atmosphere thick enough to capture stellar UV photons. Clearly, this escape rate rules out the possibility that a hydrogen-helium envelope is present as it would escape in only $\sim 1$\,Ma. A water vapor atmosphere would escape in $\sim 1$\,Ga, indicating that this is a plausible scenario.} The origin of CoRoT-7b cannot be inferred from the present observations: It may have always had a rocky composition; it may be the remnant of a Uranus-like ice giant, or a gas giant with a small core that would have been stripped of its gaseous envelope.}
{\modif{With high enough sensitivity, spectroscopic transit observations of CoRoT-7 should constrain the composition of the evaporating flow and therefore allow distinguishing between a rocky planet and a volatile-rich vapor planet. In addition, the theoretical tools developed in this study are applicable to any short-period transiting super-Earth and will be important to understanding their origins}.}


   \maketitle
%

\section{Introduction}

The newest planet discovered by space mission CoRoT is remarkably interesting. CoRoT-7b
 is not only the first super-Earth with a measured radius, but orbits extremely close to its parent star,
only 4.27 stellar radii away \citep{CoRoT-7b}. \modif{Its radius, and orbital period are $R=1.68\pm0.09\, R_{\oplus}$, 
and $P=0.854$ days respectively, the calculated age and
equilibrium temperature are $\sim 1.2-2.3$ Ga and $1800-2600$ K \citep{CoRoT-7b}
respectively, and the mass reported by radial velocity measurements is $M=4.8\pm0.8\, M_{\oplus}$ \citep{Queloz:CoRoT7b}.}

\modif{While the combination of mass ($M$) and radius ($R$) measurements alone does not yield a unique solution for the composition of a planet \citep{Valencia_ternary,Adams_Seager_Tanton:2008}, the short period of CoRoT-7b and consequently the strong irradiation on the planet, may help constrain its composition.  In this paper, we use structure models and atmospheric evaporation scenarios
to investigate the physical nature and possible origin of CoRoT-7b. 
}
\modif{
We start by considering the fate of an atmosphere (section 2), before turning to the planet's structure. Despite the intrinsic problem of degeneracy in composition, we can establish if a planet is too large to be only rocky (its radius is larger than the maximum size of a coreless magnesium-silicate planet), or even too large to be icy (its radius is larger than the maximum size of a snowball planet) given its mass.  We describe the model used to calculate the planet's structure (section 3), and show that for a subset of radius-mass combinations within the data, CoRoT-7b would in fact be too large to be composed of only refractory material. We present our results on the composition of CoRoT-7b and discuss possible evolution scenarios including as an evaporated ice or gas giant planet (section 4).  We conclude by providing arguments for the most likely scenario for CoRoT-7b.
}

The framework presented here is applicable to any transiting
close-in super-Earth. Moreover, owing to the bias of discovering short-period 
planets, we expect many such super-Earths to be discovered in the
near future with the next phases of CoRoT and Kepler's observations.

\section{Mass Loss \label{sec: mass loss}}

 Close-in planets are vulnerable to evaporation because of intense
 irradiation from their parent star.  Indeed, gas has been detected to
 flow out from the transiting gas giant HD~209458b \citep[][]{VM03}.
 Certainly, CoRoT-7b, whose mean density is \modif{$4$-$8\,\rm g\,cm^{-3}$}
 is denser than HD~209458b ($\overline\rho=0.33\,\rm g\,cm^{-3}$) by at least an
 order of magnitude. However, the UV flux received by CoRoT-7b is
 greater by an order of magnitude, because it is closer to its parent
 star and somewhat younger compared to HD~209458b
 ( ($0.017$\,AU and $1.2-2.3$\,Ga compared to $0.047$\,AU and $\sim 4$\,Ga).  One could therefore expect mass loss rates that are comparable within an order of magnitude for the
 two planets.  Consequently, given that CoRoT-7b is \modif{40 to 55} times
 less massive than HD~209458b, this mass loss may have a profound
 effect on its evolution, fate and present composition. We now attempt
 to quantify this mass loss, using simple assumptions (the precise
 modelling of atmospheric escape in this planet is a difficult task
 beyond the scope of this article).

  We model the atmospheric escape using the well-known
 expression for the extreme case of energy-limited escape,
 the validity of which has been verified for gas-giant planets
 close to their star \citep[see review by][]{YLI08}:
  \begin{equation}
    \dot{M}_{\rm esc} = \frac{3 \epsilon F_{\rm EUV}}
	{\modif{4} G \bar{\rho}_p K_{\rm tide}},
  \label{eq: escape flux}
  \end{equation}
 where $F_{\rm EUV}$ is the incident flux of the stellar EUV (extreme ultraviolet radiation),
 $\epsilon$ is the heating efficiency defined as the ratio of the net
 heating rate to the rate of stellar energy absorption, $\bar{\rho}_p$
 is the mean density of the planet, $K_{\rm tide}$ is a correction
 factor to include the Roche-lobe effect \citep{E+07,LdE04}, and $G$
 is Newton's constant.
 Of course, most of the physics is hidden in the parameter
 $\epsilon$ which is mainly controlled by the ability of the upper
 atmosphere to cool. In the case of close-in gas giants,
 detailed calculations show that H$_3^+$ plays a dominant role for the
 cooling of the upper atmosphere.
In particular,
 Eq.~(\ref{eq: escape flux}) with $\epsilon = 0.4$ and 
 HD~209458b's characteristic values yields
 $\dot{M}_{\rm esc}=4 \times 10^{10} \rm g\,s^{-1}$, which is
 to be compared to values between $3.5$ and $4.8 \times 10^{10} \rm
 g\,s^{-1}$ obtained in the literature \citep[][and references
 therein]{YLI08}.

 For planets with different atmospheric compositions, one may
 question the validity of the relation.  
  \citet{T+08} have recently simulated the escape
  of the Earth's atmosphere for different EUV irradiation levels. 
  They demonstrate that, for EUV irradiation fluxes above
  $\sim10$ times the solar value, the atmosphere is in the hydrodynamic regime,
  namely that it escapes in an energy-limited fashion
  rather than through blow-off or in a
  diffusion-limited way.
  While thermal conduction is important for
    moderate EUV fluxes, implying that mass loss then depends on
  atmospheric composition, its contribution is found to become
    negligible for strong EUV fluxes --as it is in our case--.
This can also be seen by the fact that the ratio of the EUV flux that the planet receives to a typical energy flux due to thermal conduction is $\beta\approx 10^{10-11}$, and that the ratio of a typical energy flux to a thermal conduction flux is $\zeta\approx 10^{3-5}$, provided that the thermal conduction
coefficient is of the same order of magnitude as that for
hydrogen molecules (see \citet{GM07}, for a similar
discussion about HD209458 b, and \citet{WDW81} for a precise definition of $\beta$ and $\zeta$).

  As described above, the escape efficiency is controlled by
  the radiative cooling by $\rm H_3^+$ in the case of hydrogen
  atmospheres. In the case of water-rich atmospheres, oxygen from
  dissociation of $\rm H_2O$ prevents a significant amount of $\rm
  H_3^+$ from forming, which means the efficiency might be higher
  \citep[e.g.,][]{GM07}.
  Indeed, for the case of the Earth the values of the exobase temperature and velocity of
  Fig.~8 of \citet{T+08}, one obtains the mass loss rate of the order
  of $10^9 {\rm g/s}$ in the case of the highest EUV flux (=
  $100\,{\rm erg\,s^{-1}\,cm^{-2}}$), while Eq.~(\ref{eq: escape flux})
  yields for the same conditions $\dot{M}_{\rm esc} \sim 10^9 \epsilon \, {\rm g/s}$.
 For silicate
  atmospheres, no calculations exist, 
  but we can presume that in the likely absence of species that cool
  much more efficiently than H$_3^+$, Eq.~(\ref{eq: escape flux}) 
  with $\epsilon = 0.4$ should remain
  valid within an order of magnitude --we will come back to this
  particular case in \S\ref{sec:mass loss limits}. In
    conclusion, \emph{mass loss should remain relatively large whatever the
    properties of the atmosphere (and its mean molecular weight). }

 Another important quantity controlling the escape
 flux is the flux of EUV photons emitted by the star, which is
 strongly dependent on the stellar age: 
 \begin{equation}
    F_{\rm EUV} = \alpha t_9^{-\beta} a_1^{-2}, 
    \label{eq: EUV flux}
  \end{equation}
 according to recent observations of EUV emission from young stars
 \citep{R+05}; where $t_9$ is the stellar age in Ga, $a_1$ the
 planet's orbital distance in AU, and $\alpha$ and $\beta$ are
 constants.  Their best-fit result \modif{for Sun-like stars similar to CoRoT-7} is obtained with $\alpha = 29.7 \rm\,
 erg\,s^{-1}cm^{-2}$ ($\equiv \alpha_{\rm R05}$) and $\beta = 1.23$.
 With this expression and values, one obtains that $F_{\rm EUV} = 5.0
 \times 10^4 \rm \,erg\,s^{-1}cm^{-2}$ for CoRoT-7b ($t_9 = 1.8$ and
 $a_1 = 0.017$), while $2.4 \times 10^3 \rm erg\,s^{-1}cm^{-2}$ for
 HD209458~b ($t_9 = 4.0$ and $a_1 = 0.047$).

 \modif{It can be noted that with such a high UV irradiation flux,
   CoRoT-7b may be above the purely energy-limited escape regime and
   in a regime limited by the recombination of electrons and hydrogen
   nuclei, implying $\dot{M}_{\rm esc}\propto \left( F_{\rm EUV} \right) ^{0.6}$
   \citep{MCC09}. This would imply mass loss
   rates about twice lower than estimated here with \modif{$\epsilon=0.4$} and
   eq.~(\ref{eq: escape flux}). As we are concerned with orders of
   magnitude estimates, this possibility will be ignored in the rest
   of the work.}

 Using values characteristic of CoRoT-7b in Eq.~(\ref{eq: escape
 flux}), one obtains
    \begin{equation}
      \dot{M}_{\rm esc} =  1 \times 10^{11} t_9^{-\beta} f_{\rm esc} 
                                                  \,\,\, {\rm g\,s^{-1}}
      \label{eq: escape flux 2}
    \end{equation}
 with 
    \begin{eqnarray}
      f_{\rm esc} &=& 
		 \left(\frac{\alpha}{\alpha_{\rm R05}}\right)^{}
		 \left( \frac{a_1}{\rm 0.017} \right)^{-2} 
                 \left(\frac{\epsilon}{0.4}\right)^{}
		 \left(\frac{\bar{\rho}_p}{\bar{\rho}_\oplus}\right)^{-1}
		 \left(\frac{K_{\rm tide}}{0.65}\right)^{-1}.
		 \nonumber \\ 
    \end{eqnarray}
 With the reported age of CoRoT-7b, $t_9 =$ 1.2--2.3 \citep{CoRoT-7b},
 and $\beta = 1.23$, Eq.~(\ref{eq: escape flux 2}) yields values for
 the escape rate , $\dot{M}_{\rm esc}$ = (5--10) $\times 10^{10}$ $\rm
 g\,s^{-1}$, similar to that for HD209458~b.

 To obtain the total mass lost before $t_9$, we integrate eq.~(\ref{eq: escape flux}), so that
    \begin{eqnarray}
      M_{\rm esc} &=&  0.7
                       \left(
		       \frac{ t_9^{1-\beta} - t_{9,0}^{1-\beta} }{1-\beta}
		       + t_{9,0}^{1-\beta}
		       \right) 
		       f_{\rm esc} M_\oplus ,
      \label{eq: lost mass}
    \end{eqnarray}
 where $t_{9,0}$ is the time during which the EUV flux is constant and
 taken to be 0.1.  Using eq.(5) with \modif{$\epsilon=0.4$}, assuming a planet
 density that is constant in time and CoRoT-7b's characteristic
 values, one obtains a cumulative escaped mass that goes from \modif{$\sim
 2.3\,\mearth$ for a $5.8\,\mearth$ planet to $\sim 4.5\,\mearth$ for
 a $4\,\mearth$ planet}. Hence, CoRoT-7b's present mass is in agreement
 with a planet that has not lost \modif{much} more than half its
 initial mass. \modif{However, this picture can be significantly modified when the planet's density varies greatly with time, namely, in
 the case of planets with vapor or hydrogen envelopes, as we will see
 in \S\ref{sec:evaporated}.}

 Without a detailed calculation of heating and cooling effects which
 depend on the exact composition of the escaping atmosphere, this
 should be considered only as an order of magnitude estimate. However,
 it shows that for any atmosphere to be present, it must constantly be
 resupplied and that the planet may have already lost a significant
 fraction of its mass.  \modif{On the other hand, this does not mean that
 this planet happened to be detected on its way to complete
 evaporation. By integrating eq.~(\ref{eq: escape flux 2}), one finds
 that the current state is rather stable, mainly because of the weaken EUV; complete evaporation takes
 more than 10~Ga for $\epsilon = 0.4$.}

In any case, our estimates leave room for a rather large ensemble of
possibilities concerning the global composition of the planet: it may
possess iron and rocks but also \modif{volatiles or even hydrogen and
  helium}, and the question of how much of these 
may be present arises. We attempt to address this in the following
sections.

\section{Modelling interior structure and evolution}

\subsection{Procedure}

In order to calculate the possible structure and evolution of
Earth-like planets up to ice giants and gas giants, we combine two
models. For the solid/liquid regions 
we use a 3 layer (iron/rock/ice) hydrostatic model based on Vinet
and shock equations of states; each layer is assumed to be isentropic except for the conductive thermal boundary layers at the top and bottom of the mantle \citep{Valencia_et_al:2006, Valencia_ternary}. This model reproduces the Earth's structure well and has been used previously to understand super-Earths properties.  For 
gaseous/fluid envelopes, we use a quasi-static model of interior
structure and evolution that has been extensively used to model solar
and extrasolar giant planets \citep{GM95, Gu05}. The two models are
tied by using the pressure at the bottom of the gaseous/fluid envelope
as an upper boundary condition for the calculation of the structure of
the solid/liquid interior. The temperature is \emph{not} consistently
calculated between the two models. However, this should not affect the results because thermal effects 
have a negligible impact on the properties of high-pressure iron, 
rocks and solid ices.

Our purpose is to understand possible compositions of CoRoT-7b. The
thermal evolution of such a planet is uncertain because it depends on
its composition, initial state, dynamical evolution, all of which are
unknown. It also depends on atmospheric properties and opacities, two
quantities that are difficult to estimate for a planet that probably
has a very different atmospheric composition from what has been usually considered. Fortunately, those two quantities give small impacts on our results, as described below.
\modif{Following \citet{Gu05},} evolution calculations are obtained using a
simplified atmospheric boundary condition 
\begin{equation}
T_{10}=T_{0*}\left(1+{L / L_{\rm eq}}\right)^{1/4},
\end{equation}
where $T_{10}$ is the temperature at the 10 bar level, $L$ is the
planet's intrinsic luminosity, $L_{\rm eq}$ corresponds to the stellar
luminosity that it receives and $T_{0*}$ is chosen
equal to 2500\,K to account for the presence of a greenhouse effect
similar to what is obtained for the atmospheres of close-in giant
exoplanets \citep[e.g.][]{Iro05}. In fact, because the thermal evolution
of highly irradiated planets is rapidly governed by the growth of an
inner radiative zone, it is weakly dependent on the choice of the
outer boundary condition. What is most important to us is that the
high ($\sim 2000\,$K) irradiation temperatures of CoRoT-7b maintain
the atmosphere well above the condensation temperature of water, so
that a vapor atmosphere may be present for a long time. (This is
contrary to planets in colder environments which require a large $L$
to maintain photospheric water vapor, and therefore cool quickly until they become solid). 

For opacities in gaseous envelopes, we use the
Rosseland opacity table of \citet{AF94}. The table is valid for a
hydrogen-helium solar composition mixture so that its application to
other atmospheres (e.g. one mainly formed with water vapor) may be
questioned. We point out however that the cooling is generally
controlled by the opacity in a region at a pressure $P\sim 1-10\,$kbar and
$T\sim 3000\,$K, for which the opacities are extremely uncertain,
regardless of the assumed composition \citep{Gu94}. At these pressures and
temperatures, it is mostly controlled by collision-induced absorption
by molecules in the infrared, and by the presence of electrons that
yield important absorption (e.g. from H$^-$ for a hydrogen rich gas)
at visible wavelengths. As a result, the opacities increase rapidly
with increasing $P$ and $T$, whatever the assumed composition. 
The switch from an almost isothermal external layer to
a nearly adiabatic envelope in deeper regions is expected to occur abruptly. 
In this case also, the
quantitative uncertainties on the underlying physical parameters are
large, but they have a limited impact on the result, and they do not
change qualitatively our conclusions.

Finally, the boundary condition at the bottom of the envelope is
defined as a radius provided by the hydrostatic model of the
solid/liquid interior, and a luminosity:
\begin{equation}
L_0=\dot{\epsilon}_{\rm radioactive} M_{\rm R}+ C_V {dT_{\rm Fe+R}\over
  dt} M_{\rm Fe+R},
\end{equation}
where $\dot{\epsilon}_{\rm radioactive}$ is the radioactive luminosity per
unit mass, $M_{\rm R}$ is the mass of the (rocky) mantle, $C_V$ is the rock + iron core specific heat, and $T_{\rm
  Fe+R}$ is a characteristic temperature of the solid region of mass $M_{\rm
  Fe+R}$. In the calulations, we assume a chondritic value $\dot{\epsilon}_{\rm radioactive}=2\times 10^{20}\rm\,erg\,s^{-1}\,\mearth^{-1}$, $C_V=10^7\,\rm
erg\,g^{-1}\,K^{-1}$ and $dT_{\rm Fe+R}/dt=dT_{\rm env}/dt$, where
$T_{\rm env}$ is the temperature at the bottom of the envelope. 
However, we found results to be quite insensitive to the
  choice of the inner boundary condition because for most cases, $L$
  was found to be significantly larger than $L_0$.

\subsection{States of matter inside CoRoT-7b\label{sec:states}}
\modif{We now describe the different phases and states of matter for a generic super-Earth given all possible compositions and emphasize the relevant structure for a short-period planet like CoRot-7b}. 

\subsubsection{Hydrogen and helium}

In Uranus and Neptune, hydrogen and helium form about 1 to
$4\,\mearth$ of the planets' outer envelopes \citep[e.g][and
references therein]{Gu05}. While it is not necessary expected in a
planet as small in size as CoRoT-7b, it is interesting to consider
them and provide upper limits to their abundances in the planet. 

Of course, given the temperatures to be considered ($\sim 2000$\,K and
above) and pressures well below a Mbar, hydrogen and helium are
expected to behave as a gas with hydrogen in molecular form \citep[see
phase diagram in][]{Gu05}. The equation of state considered for
modeling their behavior is that of \citet{SCVH}.

\subsubsection{Water and ``volatiles''}

\modif{
Because of their large abundances, moderately refractory species such as water, methane, ammonia are crucial building blocks of planetary systems. They are often grouped within the denomination of ``ices'' in the literature. In order to avoid the confusion with solid water, we prefer to call them ``volatiles'' and will use this term throughout the rest of the article.
}
\modif{
In a primordial disk with near-solar composition and temperatures below $\sim 200$\,K, volatiles  are by far the dominant solid species to condense \citep{Barshay_Lewis:1976}. Among those, water dominates, first because oxygen is more abundant, and second because water condenses at higher temperatures than ammonia and methane. In a solar composition mixture, oxygen is more abundant than carbon by a factor 1.8, to nitrogen by a factor 7.2 and to magnesium, silicium and iron by factors 12, 15 and 15, respectively \citep{Asplund:CompoSun}. We hereafter use water as a proxy for volatiles in general, an assumption that is minor compared to other sources of uncertainty. 
}

\begin{figure}[htbp]
\begin{centering}
\includegraphics[angle=0,width=0.47\textwidth]{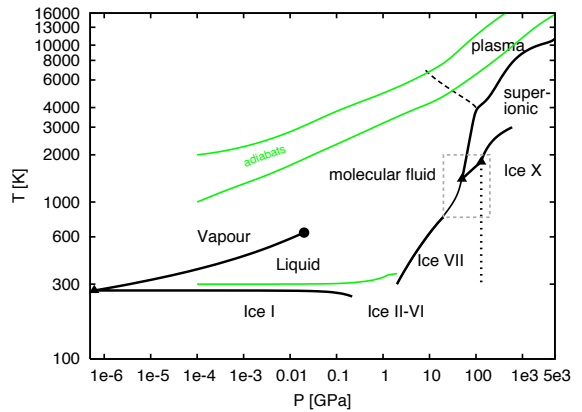}
\end{centering}
\caption{Phase diagram of water from 0.01 bar to 50 Mbar and 300 to 16000 K. \textit{Black solid lines}: phase-transition boundaries, \textit{Triangles}: Triple points, \textit{Circle}: critical point; \textit{grey solid lines}: adiabats starting from 300, 1000, and respectively 2000 K at 1 bar; \textit{black dashed line}: continous transition from molecular dissociated water to water plasma; \textit{grey dashed box}: region of high uncertainty and contradictive experimental and theoretical results. For $P>4$ GPa and $T\geq 1000$ this diagram relies on data from FT-DFT-MD calculations \cite{French+09}. The melting curve is taken from \cite{FW06-ice1}, the saturation curve from \cite{WP02-liq}, and the phase boundary of Ice~VII is adapted from \cite{Goncharov+09} for pressures lower than 30~GPa. For higher pressures, experimental and theoretical investigations predict different  phase boundaries Ice~VII$-$molecular water, Ice~VII$-$superionic water, Ice~VII$-$Ice-X (controversial as a hard boundary from experiments \citep{Hemley:1987}), IceX$-$superionic water and different locations of corresponding triple points~\cite{French+09,Goncharov+09,Schwager+04}.}
\label{fig:water}
\end{figure}

From the phase diagram of H$_2$O (Fig.~\ref{fig:water}) it is clear that with an \modif{atmospheric} temperature above 1000\,K, the planet would be composed of supercritical water. If the planet follows an adiabat, it will remain in vapor form transforming eventually into a plasma. If instead, the planet had a surface temperature below the melting point of water (e.g. because it formed far from the central star), it would exhibit different high pressures forms of ice, up to a regime where ice VII and ice X (for the massive icy planets) dominate.

The EOS used for water vapor is obtained from a combination of data obtained from a finite temperature molecular dynamics (FT-DFT-MD) simulation by French et al. (2009) and of the Sesame 7150 EOS (see Kerley 1972). The FT-DFT-MD data are used for $T=1000-10,000$\,K and $\rho=2-7\rm\,g\,cm^{-1}$ as well as for $T=10,000-40,000$\,K and $\rho=5-15\rm\,g\,cm^{-1}$. Sesame 7150 data are used elsewhere. The two EOS are joined by interpolation of isotherms.

\subsubsection{Silicates}
\begin{figure}[htbp]
\begin{centering}
\includegraphics[width=0.47\textwidth]{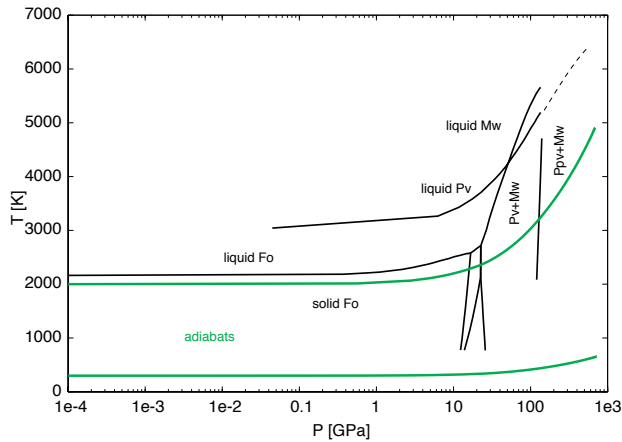}
\end{centering}
\caption{Simplified $P-T$ phase diagram for relevant silicates on super-Earths. Data is taken from \citet{Presnall:1995} for the Mg-silicate end member, with phase boundaries from solid forsterite (dominant in the upper mantle), to Earth's lower mantle materials, perovskite (pv) and magnesiowustite (mw). The melting curve for pv and mw, which remains controversial, are shown, as well as an extrapolation of the pv's melting curve to higher pressures. The phase boundary of post-perovksite (ppv) was calculated from \citet{Tsuchiya:2004}.   Adiabats at 300 K and 2000 K and 1 bar are shown for reference.} 
 \label{fig:silicates}
\end{figure}

Although silicates are basically made of (Mg,Fe) O + SiO$_2$, the phase diagram relevant for the mantle is very complicated due to the different minerals that can be formed and the presence of iron and other minor elements (Ca, Al). 
We show the relevant phases for the magnesium end member in
Fig.~\ref{fig:silicates}.  The diagram shows the forsterite
(Fo: Mg$_2$SiO$_4$), perovskite/post-perovskite (pv/ppv: MgSiO$_3$),
magnesiowustite (mw: MgO) system. In addition, the upper mantle would also include the pyroxene phases (Mg$_2$Si$_2$O$_6$)).

We show two adiabats calculated at 300 K and 2000 K and 1
bar for comparison. Both melting curves of pv and
mw show a steep slope that can pose a barrier to melting of the
interior. Given that we do not know the melting behaviour of
post-perovskite or of MgO at large pressures, it
remains unclear if the lower-most mantle of super-Earths can easily
melt or not. 

It should be noted that melting will depend on the amount of iron in
  the mantle (i.e. the magnesium number), but also on the abundance of
  minor species, something not included in Fig.~\ref{fig:silicates}.
  As an example, on Earth, decompression melting can occur at temperatures around 1300K \citep{Hirschmann:2000}. 
We do not attempt to determine the
  fraction of the planet's surface that may be molten, but note that
  it may be relatively large.

\subsubsection{Iron}
\begin{figure}[htbp]
\includegraphics[width=0.47\textwidth]{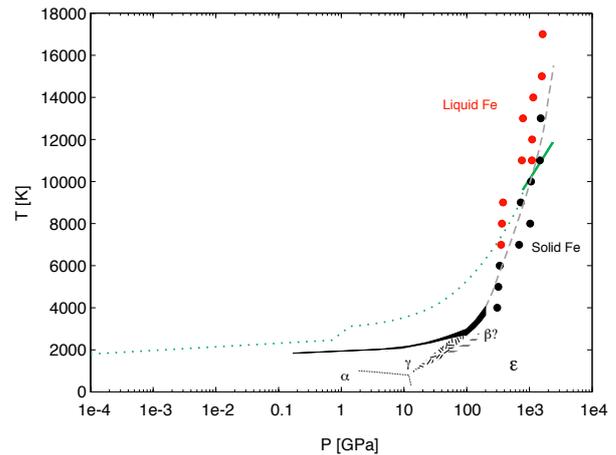}
\caption{$P-T$ phase diagram for iron. Values for pressures below 200
  GPa were adapted from \citet{Boehler:2000} and references therein.
  The black region shows the agreement in the melting curve of iron at
  relatively low pressures. Data points for melting in the high
  pressure regime of 306-1625 GPa are from \citet{Morard:2009} .  Red
  points correspond to the liquid phase, while black points are solid
  Fe. The dash curve is a melting line drawn to approximate
  the boundary suggested by the results from the ab-initio calculations. \modif{ The temperature profile for an Earth-like CoRoT-7b is shown in green. The dotted part corresponds to the mantle, whereas the solid line corresponds to the core's temperature}. 
\label{fig:iron}}
\end{figure}

modif{While forming a pure iron planet is very unlikely, evaporating the mantle of a Mercury-like planet might be possible. We show the phase diagram for iron in Fig.~\ref{fig:iron}. Different phases of iron have been identified in the low pressure regime \citep{Boehler:2000} with relative agreement. The $\epsilon$ phase seems to be the most relevant to Earth's core. A pure-iron planet might transition between different phases of iron depending on the pressure-temperature profile.
}
\modif{
However, the high pressure regime in which most of the cores of super-Earths would be in, is still inaccessible to experiments. Thus, it is unclear if there are any other higher pressure phases of iron unidentified at this point. One study \citep{Morard:2009} has reported from ab-initio calculations the melting behaviour of iron in the tens of megabars pressure regime (red and black symbols in Fig.~\ref{fig:iron}). The melting boundary is quite steep implying that pure-iron planets are likely to be mostly solid. However, planets with mantles have hotter interiors due to their insulating character. 
}

\section{Inferring Composition}

\subsection{CoRoT-7b as a rocky (iron+rock) planet}
\subsubsection{Description}

\modif{We first explore the case in which the planet is of telluric
composition. This implies a variety of compositions, from a pure magnesium-silicate planet (with no iron) to a pure iron planet. The former would yield the largest size for a rocky body, while the latter would be the smallest. Either case is unlikely. During the cooling of a protoplanetary disk, iron and silicates are condensed out at similar temperatures so that if iron is present,
so are silicates, and visceversa, especially in large objects. Furthermore, the variety in structure for rocky planets includes those that are differentiated and undifferentiated. The former has a layered structure with the core composed mainly of iron, and in the case of Earth some nickel and a light alloy \citep{McDonough_Sun:1995} below a silicate mantle.  The mantle can also incorporate iron within the oxide structure replacing the magnesium site. Undifferentiated planets would have all of their iron content embedded in the mantle rocks.  The amount of iron with respect to magnesium in the mantle (the magnesium number) speaks to the degree of differentiation of a planet and is a consequence of early formation, when
the part of Fe that remained immiscible differentiated to form the
core. The iron content x$_{\rm{Fe}}=$Fe/(Mg+Fe) for Earth has been estimated at 0.1 \citep{McDonough_Sun:1995}, while for Mars it is calculated to be 0.20-0.25 \citep{Ohtani:MarsComp}. For super-Earths this number may greatly vary, although due to larger accretional energies, bigger planets may be expected to be differentiated.}

\modif{On the other hand, the composition of planets can be compared by looking at bulk elemental ratios such as Fe/Si. For Earth this number is considered to agree with that of CI chondrites \citep{McDonough_Sun:1995} and is $\sim$ 2. Although it is unclear if the planets should have the same Fe/Si bulk ratio as their host star, it is a reasonable assumption. Mercury is an anomaly in the solar system. However its anomalously high iron content may be related to secondary formation processes like giant impacts and erosion, which may have dramatic effects on planetary compositions.}  

\begin{figure}[htbp]
\begin{centering}
\includegraphics[width=0.47\textwidth]{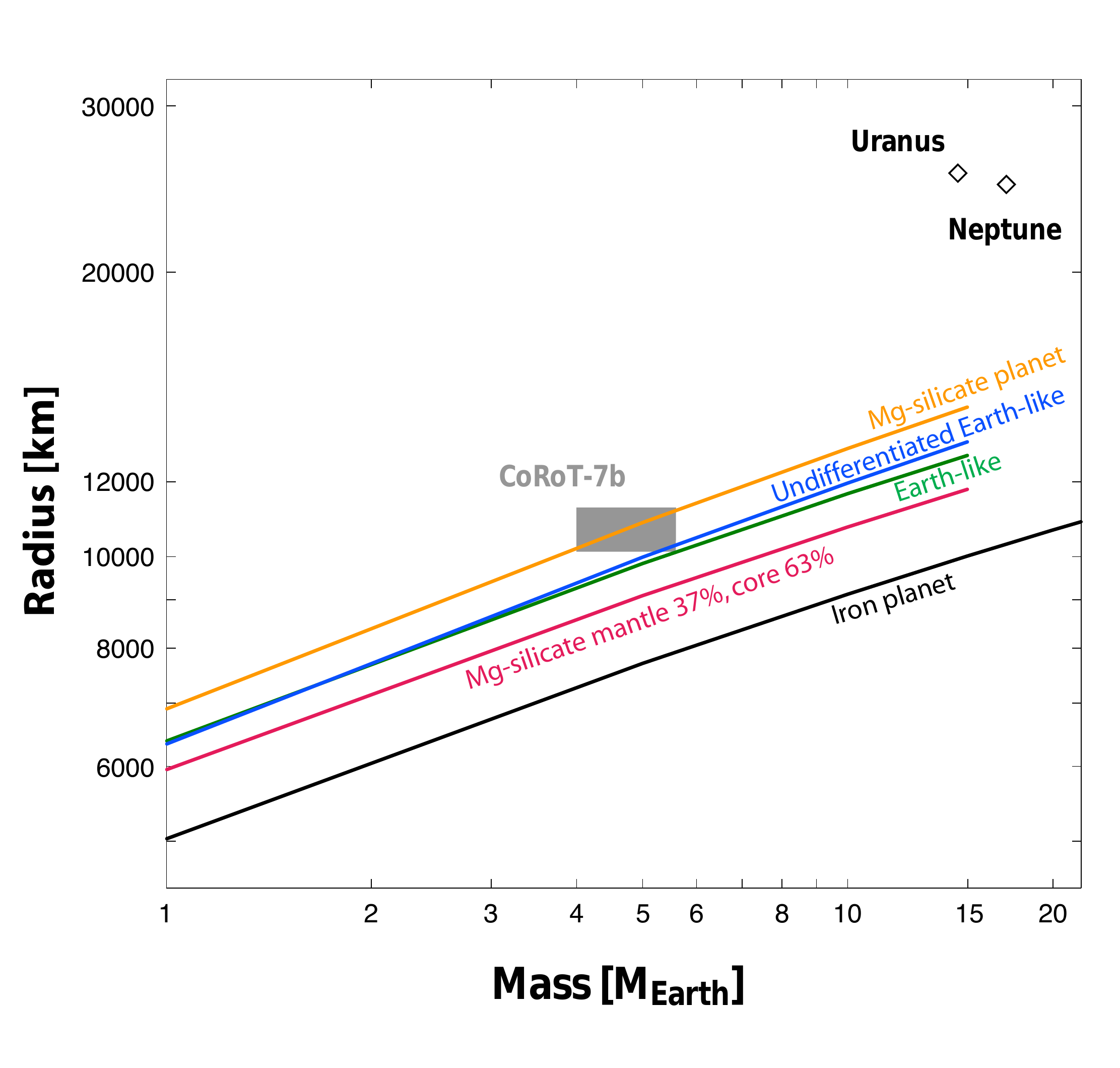}
\end{centering}
\caption{\modif{Mass-radius relations for planets made of iron and rocks. The compositions considered are pure iron, 63\% core and 37\% mantle, Earth-like (33\% core and 67\% mantle with $x_{\rm{Fe}}$=0.1 iron content), an undifferentiated planet with bulk Fe/Si=2 (Earth-like), and 100\% Mg-silicate mantle planet. The surface temperature is taken to be  1800 K. The shaded area corresponds to CoRoT-7b's measured radius and mass. Any combination of M-R that lies above the relation for pure Mg-silicate planets (orange line), requires a composition that includes volatiles. Compositions with larger amounts of iron would progressively lie below this line.} \label{fig:M-R}}
\end{figure}

\modif{To infer CoRoT-7b's composition we considered different possibilities:   1) A pure Mg-silicate planet; 2) an Earth analog (i.e. a differentiated planet with $x_{\rm{Fe}}=0.1$, and a core that is 33\% by mass); 3) an undifferentiated planet with the same bulk Fe/Si ratio as Earth's, which we obtain with an iron content of $x_{\rm{Fe}}=0.76$ by mol; (4) a planet with no iron in the mantle and a core-mass fraction of 63\% (i.e a super-Mercury); and (5) a pure iron planet. The mass-radius relations are shown in Figure \ref{fig:M-R}.}

\modif{To calculate the Fe/Si ratio of the differentiated and undifferentiated planets we considered a mantle composed of $\frac{1}{2}$(Mg$_{(1-x_{\rm{Fe}})}$,Fe$_{x_{\rm{Fe}}}$)$_2$SiO$_4$ + $\frac{1}{2}$(Mg$_{(1-x_{\rm{Fe}})}$,Fe$_{x_{Fe}}$)$_2$Si$_2$O$_6$ in the upper mantle and $\frac{3}{4}$(Mg$_{(1-x_{\rm{Fe}})}$,Fe$_{x_{\rm{Fe}}}$)SiO$_3$ + $\frac{1}{4}$(Mg$_{(1-x_{\rm{Fe}})}$,Fe$_{x_{\rm{Fe}}}$)O, in the lower mantle and lowermost mantle (the post-perovskite region). In addition, we used a Ni/Fe ratio of 17 and had a light alloy in the core of 8\% by mass after Earth's composition \citep{McDonough_Sun:1995}.
}

\modif{
Owing to the fact that the largest radius for a rocky CoRoT-7b
  corresponds to the Mg-silicate planet, any radius above this line
  reveals the presence of volatiles. Rocky planets with increasing
  amounts of iron content, whether differentiated or undifferentiated
  will have a mass-radius relation lying progressively below the pure
  Mg-silicate line. Coincidentally, this 'super-Moon' composition conforms to the
  smallest and largest mass of CoRoT-7b.  However, it is
  unrealistic as iron is expected to be present in some amount, so
  that the lower end of CoRoT-7b's mass range can not be justified
  without the presence of volatiles. Massive planets mostly made of
  silicates and with very little
  iron (``super-Moons'') are unlikely to
  exist: it is difficult to imagine how the special conditions that
  led to the formation of our Moon could also prevail for a planet 500
  times more massive, and with the dissapearance of the massive
  iron-rich counterpart object (equivalent to the Earth). 
}

\modif{The difference between differentiated and undifferentiated planets is that the latter are slightly larger and this effect becomes more noticeable with increasing mass. Our result agrees with that of \citet{Elkins_Seager:2008}. The difference in radius for planets with 1, 5, 10 and 15 $M_\oplus$ is of 0.7\%, 1.5\%, 2.6\% and 3.4\% respectively.  Thus, from mass and radius measurements it seems implausible to distinguish between a differentiated and undifferentiated planet. However, perhaps atmospheric evaporation of silicates indicating the amount of iron might help infer the state of differentiation. 
}

\begin{figure}[htbp]
\begin{centering}
\includegraphics[width=0.3\textheight]{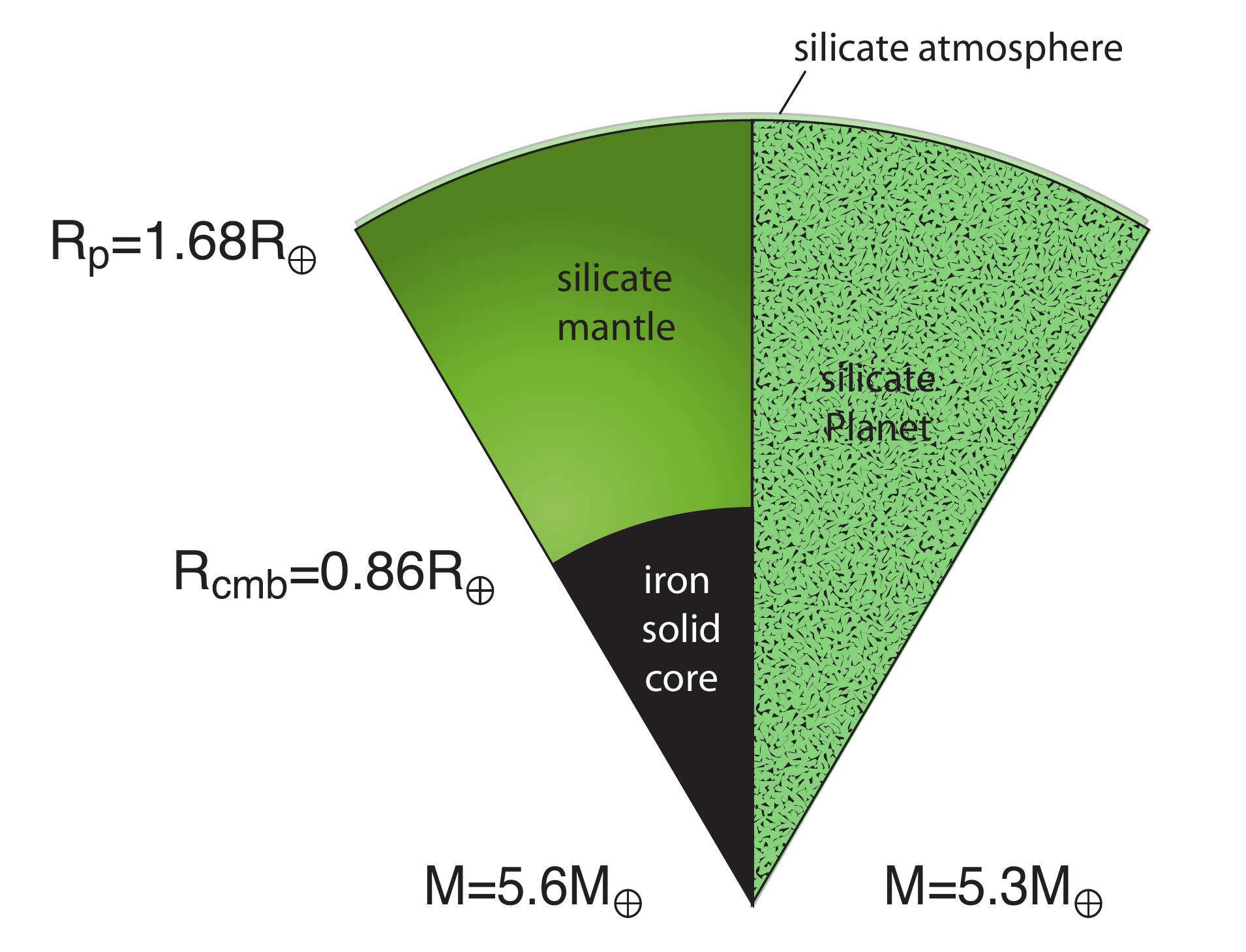}
\end{centering}
\caption{\modif{Drawing to show the interior structure envisioned for a
  rocky planet. Two scenarios with Fe/Si = 2: (left)
Earth-like (10\% by mol of iron in the mantle, and core-mass fraction of 33\%), or (right) undifferentiated}. \label{fig:Fe+R}}
\end{figure}

\modif{We exemplify the different structures of an Earth-like composition and the equivalent undifferentiated planet in terms of the Fe/Si. For a fixed radius of $R=1.68\,R_\oplus$ the differentiated and undifferentiated cases would have a mass of 5.6 and 5.3 respectively. Figure \ref{fig:Fe+R} depicts their different interiors. }

\modif{Furthermore, the state of the core is shown in Fig. \ref{fig:iron}, where we show the temperature profile for an Earth-like composition including concentration of radioactive sources and an age of 1.8 Gy. The core's temperature lies right at the melting boundary, so that given the uncertainties in temperature structure there is a possibility of an outer liquid core. These uncertainties include the exact concentration of radioactive sources (perhaps more potassium), the existence of a boundary layer right above the core, and even possibility of a layered mantle due to the different viscosities between perovskite and postperovskite. 
} 
\modif{
Given the very steep behavior of the iron solidus, it seems unlikely that massive terrestrial planets could have liquid cores, especially those with a surface temperature that allows for liquid water. 
}

\subsubsection{The atmosphere}
Because of the large irradiation the planet is subject to, its surface may
be heated to extremely high temperatures. Given the star's
characterisics ($T_{\rm eff}=5275$\,K, $R_\star=0.87\,\rm R_\odot$)
and planet's semi-major axis ($a=4.27 R_\star$), the effective
equilibrium temperature at the planet's substellar point is (assuming
a zero albedo) $T_{\rm eq}^{\rm sub}=2570$\,K. Distributed evenly on
the planet's surface, this temperature is $T_{\rm eq}^{\rm
  global}=1820$\,K. Even though the exact temperature depends on the
emissivity and albedo, it is clear that silicates (or iron if present
at the surface) should be molten by the intense heat, at least on part
of the planet for temperatures above 2000 K (see Figs. \ref{fig:silicates} and \ref{fig:iron}). 

A first consequence of a molten surface is that volatiles should be
efficiently outgassed from the planetary interior
\citep[e.g.][]{SF07}. However, because the total mass of those
volatiles would be as small as at most a few 10~\% of the planet's mass,
this is a temporary effect and given the significant mass loss, the
massive atmosphere thus formed should disappear quickly (on a
timescale of $10^6-10^8$\,yrs, based on the arguments in \S
\ref{sec: mass loss}) . The remaining planet should then contain only
refractory material, with vapor in equilibrium with the lava, and an
evolving composition as a function of the mass loss.

Specific models for the chemistry of the atmosphere of
  evaporating silicate super-Earths \citep{SF09} indicate that a
  planet such as CoRoT-7b with a composition similar to that of the
  bulk silicate Earth would have an equilibrium atmosphere with a
  pressure initially between $10^{-6}$ and $10^{-2}$ bars
  (corresponding to our range of extreme temperatures).  The
  evaporating vapor atmosphere should be mainly composed of Na, then
  SiO, O and O$_2$, then Mg, as the less refractory species are
  progressively lost.  \citet{SF09} find that the pressures decrease
  by about 1 order of magnitude when Na is lost,  and then by a
  further 2 to 3  orders of magnitude at 90\% of total erosion. 

With such a thin atmosphere, the planetary radius measured from the
transits can be considered as that of the solid/liquid surface of the
planet.

\subsubsection{Limits to the mass loss?\label{sec:mass loss limits}}

Compared to the arguments presented in \S\ref{sec: mass loss}, the
planet's erosion may be reduced if (i) UV photons are not fully
absorbed in the atmosphere but hit the surface of the planet, or (ii)
the ``supply'' of atmosphere is slowed by the need to deliver heat to
pass the latent heat barrier.

The photoionization cross-section of atomic species such as H, O, Fe,
Mg, Si is between $\sigma_{\rm UV}\approx 10^{-19}$ and
$10^{-17}\rm\,cm^2$ in the 10-50 eV energy range \citep{V+96}. The
unit optical depth for UV photons corresponds to a pressure $P_{\rm
  UV}=\mu m_{\rm p} g/\sigma_{\rm UV}$, where $\mu$ is the mean
molecular weight of the atmosphere, $m_{\rm p}$ the proton's mass and
$g$ the planet's gravity. Using $g\approx 1000\rm\,cm\,s^{-2}$ and
$\mu\approx 10$\,g one finds $P_{\rm UV}\approx 10^{-7}$ to
 $10^{-9}$\,bar. (For comparison, $P_{\rm UV}\approx 10^{-9}$\,bar for
  gas giants -- see \citet{MCC09}). 

Because those values of $P_{\rm UV}$ are much smaller than
those of the vapor pressure estimated above, it appears that the
equilibrium vapor atmosphere is able to efficiently absorb stellar UV
photons that will drive the escape from the planet. Note however that
the precise rate of escape depends on cooling processes and
hydrodynamical modeling much beyond the scope of this work.

Let us now consider whether a bottleneck to the escape may be caused
by the need to vaporize material that is initially solid or liquid. In
order to do so, we balance the energy required for the sublimation at
a rate $\dot{M}_{\rm sub}$ with the absorbed heat flux $\pi R_p^2
F_\star$, where $F_\star$ is the stellar irradiation. Note that
$F_\star$ is formally the stellar flux that reaches the ground, but
given the thinness of the atmosphere, this is equivalent as the
irradiation flux at the top of the atmosphere.  The sublimation rate
can hence be written:
\begin{equation}
\dot{M}_{\rm sub}={\pi R_p^2 F_\star \over \cal{L}_{\rm sub}},
\label{eq:Msub}
\end{equation}
where ${\cal L}_{\rm sub}$ is the latent heat of sublimation.  Using
${\cal L}_{\rm sub}\approx 10^9-10^{10}\rm\,erg\,g^{-1}\,K^{-1}$
(typical for iron and silicates), we
find $\dot{M}_{\rm sub}\approx 10^{17}$ to $10^{18}\rm\,g\,s^{-1}$,
i.e. at least 6 orders of magnitude larger than the mass loss we previously
derived. Hence, there is no mass loss suppression due to latent heat effects. 

We thus conclude that at this orbital distance, less than 5 stellar
radii away from its star, the planet should be eroding even if made of the
most refractory materials!

\subsubsection{Evaporation of a rocky CoRoT-7b \label{sec:evapo-rocky}}

\begin{figure}[htbp]
\begin{centering}
\includegraphics[width=0.47\textwidth]{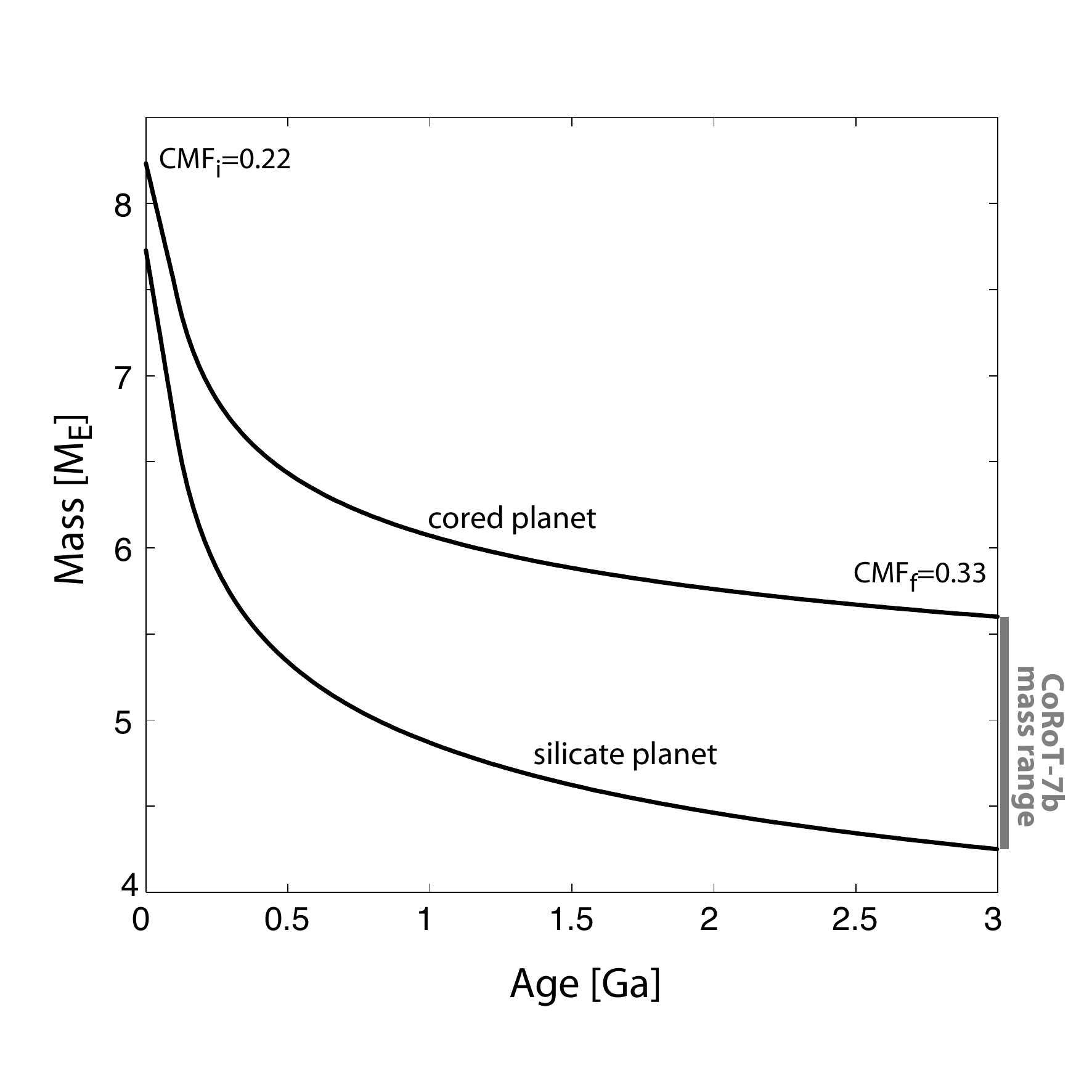}
\end{centering}
\caption{Evolution in mass for CoRoT-7b for a rocky composition. \modif{Two present compositions where considered: a 'super-Moon' (little or no iron content), and an Earth-like composition (33\% iron core, 67\% silicate mantle), corresponding to the lightest and densest cases for a rocky CoRoT-7b, respectively.  The initial and final proportions of core mass fraction are shown. The mass loss is more significant early on due to its inverse relation to density. The mass loss is calculated from Eq.~\ref{eq: escape flux 2}, with $\epsilon=0.4$.}  \label{fig:massloss-sil}}
\end{figure}

\modif{We now investigate the possible precursors of CoRoT-7b given that its mantle is being subjected to considerable erosion. Calculations of this erosion depend on the orbital evolution of the planet, the decrease in stellar EUV flux, but mostly on the change in bulk density as layers (or perhaps selective components) of the planet are stripped away. In this section we are mostly concern with the effect of changing density}. We integrate Eq.~\ref{eq: escape flux} backward in time to assess the mass loss experienced and calculate the composition of the precursors. We use the internal model to obtain the planet's structure at each time step and calculate its density. \modif{The chosen efficiency for this calculation is $\epsilon=0.4$. We consider two present compositions: a 'super-Moon' and an Earth-like composition. They correspond to the lightest and densest compositions for a rocky CoRoT-7b. The results are shown in Fig.~\ref{fig:massloss-sil} and exemplify the role of changing density. }

\modif{We find that small planets suffer a larger atmospheric loss, due to the inverse dependence on average density. Also, as planets lose their silicate mantles, their average density either increases (for small planets) or stays relatively constant (for massive planets) owing to a cancelling effect between a reduction in size and mass. Thus, the rate of atmospheric loss decreases through time. Assuming $\epsilon=0.4$, we calculate the amount of mass lost to be 3-4 $M_\oplus$, so that if formed rocky, CoRoT-7b was initially a $\sim 8\,M_\oplus$-planet with a core that was at most 22\% by mass and Fe/Si=1.28. This is probably an upper limit, given that studies of the hydrodynamic loss of a silicate-rich atmosphere have never been done and may lead to a lower $\epsilon$ value, we do not consider probable inward migration, or uncertainties in the EUV flux.}

\subsection{CoRoT-7b as a vapor planet}

\subsubsection{Description and possible origin}
\modif{
Given CoRoT-7b's relatively large mass, it should have
  originally accreted a significant mass of gases, i.e. hydrogen
  and helium (e.g. \citet{IG06}), and volatiles (a.k.a
  ``ices'') 
  (e.g. \citet{McNeil_Nelson:SEForm}). Given the present
 extreme stellar irradiation, one would expect volatiles, if
  present on the planet, to be in vapor form (see
  \S\ref{sec:states}). However, we first examine the alternative
  hypothesis that if the planet migrated from large orbital distances
  \citep[e.g.][]{L+96}, water had time to cool and solidify.
}

\begin{figure}[htbp]
\begin{centering}
\includegraphics[width=0.3\textheight]{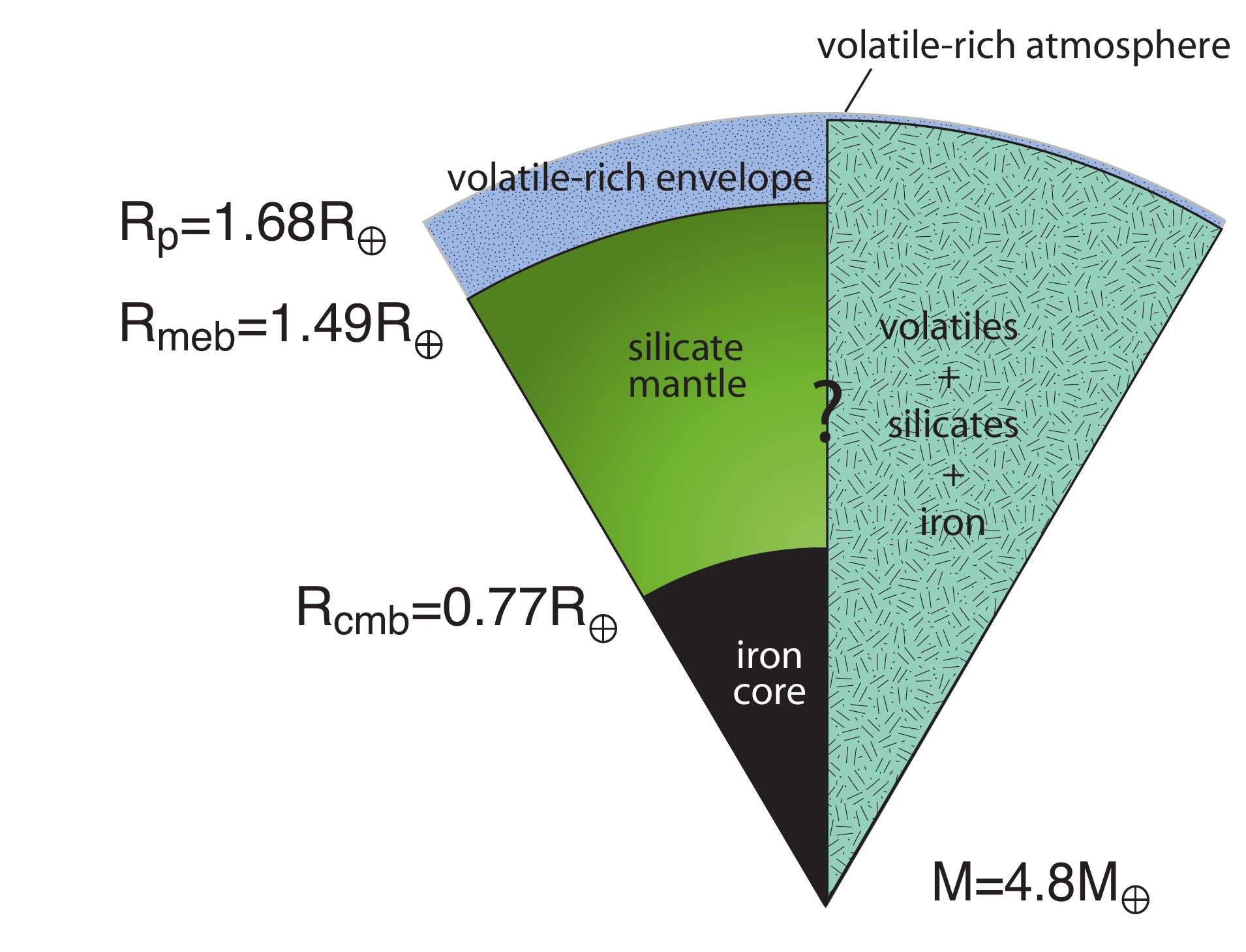}
\end{centering}
\caption{Drawing to show the interior structure envisioned for a
  planet made of iron, rocks and volatiles in vapor form (possibly with H-He)\label{fig:Fe+R+steam}}
\end{figure}

Let us consider a planet of $M_{\rm p}=10\,\mearth$ made mostly of water (for
simplicity). The gravitational energy transformed into internal energy
during its formation is $E_{\rm i}\approx 3/10\ GM_{\rm p}^2 /R_{\rm
  p}$. In the absence of irradiation (if far enough from the star),
the planet is initially made of vapor. If vapor is present in the
photosphere (i.e. without an optically thick layer of hydrogen
and helium), it will maintain a high atmospheric temperature
(and therefore a rapid cooling) until complete condensation of water
onto the interior is reached. Given an effective temperature
  (as obtained from the temperature at which the saturated vapor
  pressure \citep[e.g.][]{Em94} is equal to the photospheric pressure, $P_{\rm
    vap}(T)=2/3\,g/\kappa$), an upper limit to the time to
solidification is $\tau_{\rm solid}=E_{\rm
  i}/(4\pi R_{\rm p}^2\sigma T_{\rm eff}^4)$, i.e. for $R_{\rm
  p}\approx 15,000\rm\,km$, $\tau_{\rm solid}\approx 100$\,Ma when
using a low opacity $\kappa=10^{-2}\,\rm cm^2\,g^{-1}$ ($T_{\rm
  eff}\approx 320$\,K). On the other hand,  if ice grains and/or water droplet in the
atmosphere prevent it from cooling efficiently, $\kappa\approx 10^{2}\,\rm
cm^2\,g^{-1}$ ($T_{\rm eff}\approx 210$\,K) so that $\tau_{\rm
  solid}\approx 1$\,Ga. 

At least in theory, there is thus a possibility that CoRoT-7b is a
planet that was formed at large distances from the star, had
sufficient time (hundreds of millions of years) to solidify before
being brought to an orbital distance of 0.017\,AU, where ice would be
sublimating again.

\modif{
However, we believe that this is very unlikely for two reasons: First,
a planet that massive should accrete some hydrogen and helium
\citep{IG06}, and even only a few bars of these species would mean
that water would still be in vapor form in the interior at smaller
effective temperatures than estimated above in the pure water case (a
situation similar to that of Uranus and Neptune). This would imply a
(much) slower cooling, and thus retaining water in vapor form for a
longer time. Note that invoking a putative evaporation of an outer
hydrogen atmosphere would require an increased irradiation which would
also be unfavorable to the rapid cooling of the planet. Second,
although possibilities of slow/delayed migration exist
\citep[e.g.][]{WM03}, most of the migration scenarios require ``help'' from the
protoplanetary disk, and therefore a migration in the first millions
of years (e.g. \citet{Moorhead:2005}).
}
\modif{
In what follows, we therefore only consider the possibility that ices
are in vapor form. Figure \ref{fig:Fe+R+steam} depicts the two
possibilities that we envison: a fully differentiated planet with
iron, silicates and an extended envelope of vapor, or a fully homogeneous
planet in which iron, silicates and volatiles (in vapor form) are
thoroughly mixed. 
}


\subsubsection{Constraints on the presence of vapor}

To estimate possible amounts of vapor compatible with the
measurements of CoRoT-7b, we proceed as follows: First, for
simplicity, we only consider the case of a solid/liquid interior that
is ``Earth-like'' in composition (33\% iron core, 67\% mantle rock)
and surrounded by an envelope of \modif{volatiles in vapor form}. Our calculations of interior models for the iron+rock part as a
function of its mass $M_{\rm Fe+R}$ and outside pressure $P_0$ are found to yield radii of
order:

\modif{
\begin{eqnarray}
R_{\rm Fe+R}&=&9800\,{\rm km}\ \left(M_{\rm Fe+R}\over
  5\,\mearth\right)^{0.28+0.02\sqrt{M/5\,\mearth}}  \nonumber \\
& & \times 10^{-\left[\log_{10}(1+ P_0 /\sqrt{M/5\,\mearth})/7\right]^3}
\end{eqnarray}
where $P_0$ is in GPa units. The relation is an approximation found to be accurate
to $\pm 0.5\%$ for $M_{\rm Fe+R}$ between $1$ and $15\,\mearth$ and
$P_0$ up to $10^3$\,GPa ($=100$\,Mbar). The approximation is accurate to
$\pm 3\%$ to $10^4$\,GPa. 
}

The size of the planet with vapor is found by calculating the
evolution of an initially adiabatic planet with a specific entropy
equal to that of vapor at $10$\,bar and $2500$\,K. 
This initial state
is chosen as representative of any ``hot start'', since any evolution
from still higher entropies would have been fast. We neglect any
possible orbital evolution of the planet.

The evolution is characterized by the rapid growth of a
radiative zone just below the atmospheric boundary, similarly to what
is obtained for giant exoplanets\citep{Gu05}. This zone quickly
becomes isothermal and extends down to pressures around 10\,kbar, and
temperatures $\sim 3000$\,K. At those pressures and temperatures, the
rapid rise in radiative opacities implies that any further
extension of the radiative region must wait for a large reduction of
the intrinsic luminosity, implying a slow cooling. This implies that
results should be relatively robust, in regard of uncertainties in initial
state, opacities, age...etc. 

\begin{figure}[htbp]
\begin{centering}
\includegraphics[width=0.47\textwidth]{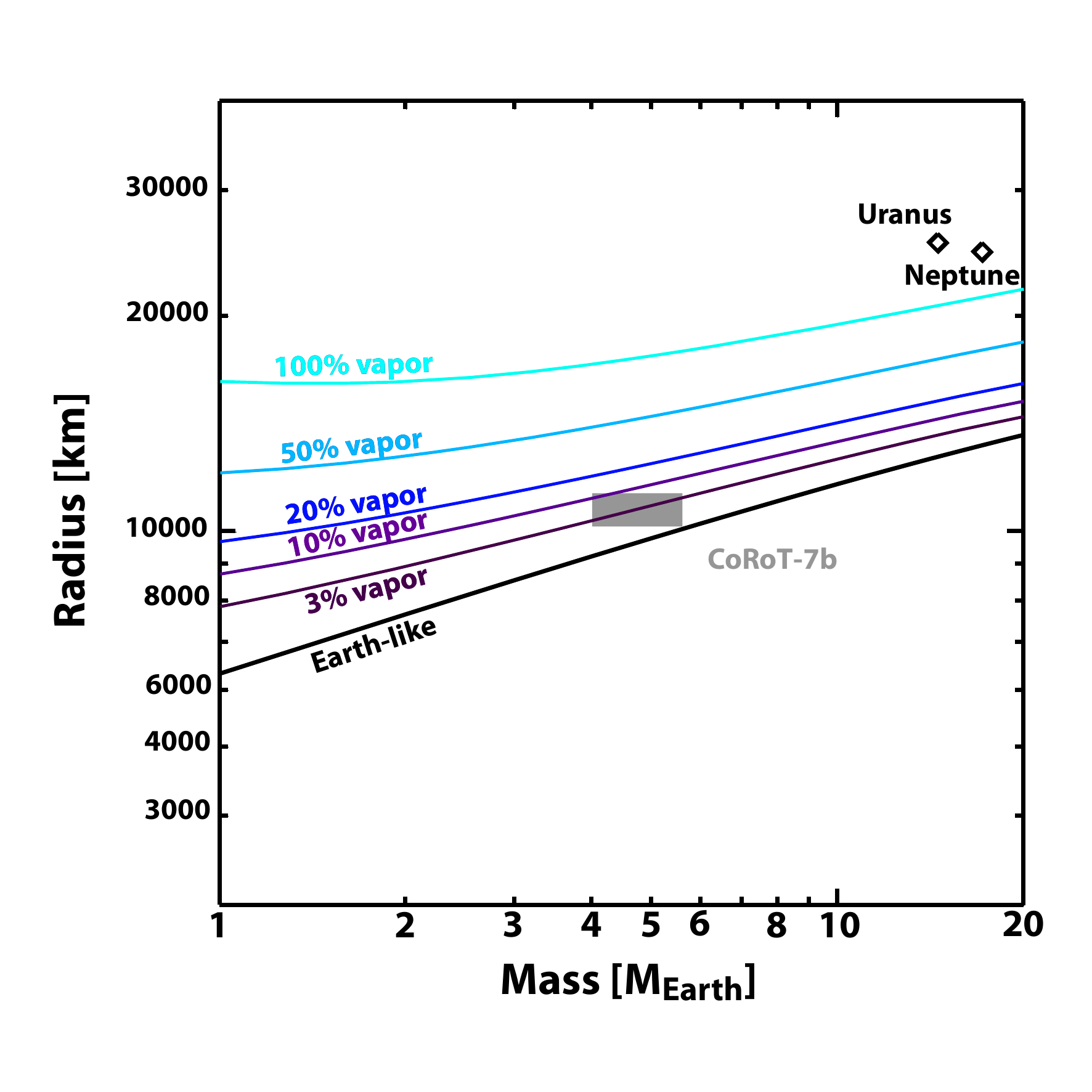}
\end{centering}
\caption{Radius as a function of mass for a planet made of iron, rocks
  and vapor. \modif{The ``Earth-like'' line corresponds to the
  limiting value of a solid Earth-like planet with a mass fraction of
  iron to rocks of 33\% and without vapor. Other lines corresponds to
  radii for planets with an Earth-like interior and a vapor (H$_2$O)
  envelope, with a ratio of the mass of the vapor envelope to the total
  planetary mass between 3\% and 100\%, as labeled. The models with
vapor have been evolved for 1\,Ga using our 
  fiducial opacity table (see text). Radii correspond to the 10\,bar level. \label{fig:M-R steam}}}
\end{figure}

Figure~\ref{fig:M-R steam} shows the resulting planetary radii after
2\,Ga of evolution for various mass fractions of vapor in the
planet. The presence of an atmosphere of vapor is found to affect the
structure and size of the planet significantly. 
We find that the upper limit on the amount of vapor
present in CoRoT-7b is \modif{$\sim10\%$, i.e. the equivalent to about
  $0.5\,\mearth$. Because the cooling and contraction occurs
    rapidly, we find that this value is robust and doesn't change by
    more than a few percent when considering cooling times between 0.1
  and 10\,Ga. Given the 
evaporation rate calculated in \S~\ref{sec: mass loss}, this implies that such an
atmosphere would last for another Ga or so. This is hence a
reasonable possibility. Our best vapor-planet model for CoRoT-7b with a total mass of
$4.8\,\mearth$, has a vapor envelope that is 3\% of the total mass and 12\% of the total
radius (see Fig.~\ref{fig:Fe+R+steam}). For this model, the envelope
is close to being isothermal: the transition between the vapor
envelope and the silicate mantle is at a temperature of $2900\,K$ for
a pressure of $14\,$GPa.}

\subsubsection{Constraints on the presence of hydrogen and helium}

\begin{figure}[htbp]
\begin{centering}
\includegraphics[width=0.47\textwidth]{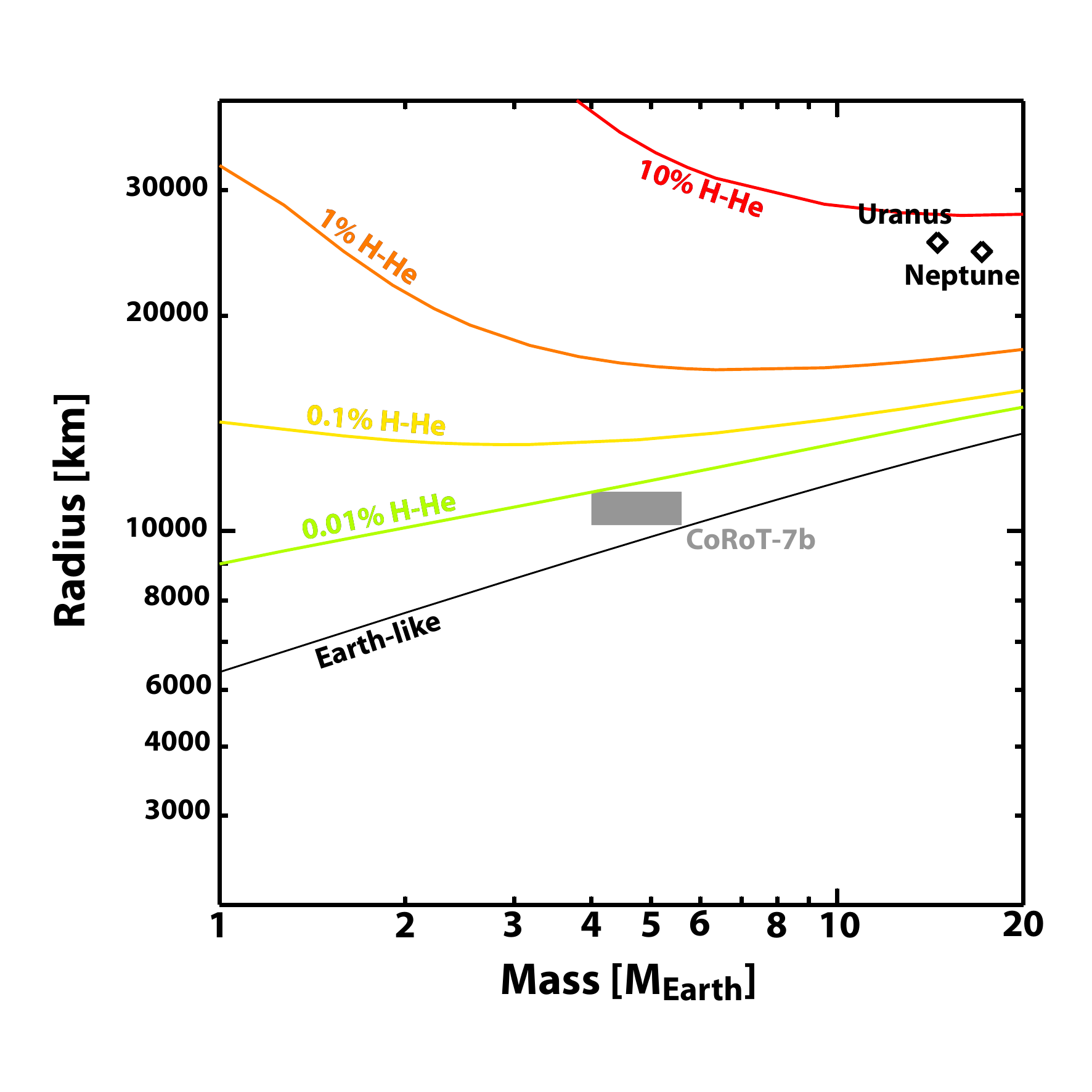}
\end{centering}
\caption{Radius as a function of mass for a planet made of iron, rocks
  and hydrogen and helium in solar proportions. \modif{The ``Earth-like'' line corresponds to the
  limiting value of a planet without vapor. Other lines correspond to
  planets with hydrogen and helium envelopes having total mass fractions between 0.01\% and
  100\% (planet made only of hydrogen and helium), as labelled. (See
  fig.~\ref{fig:M-R steam})}. \label{fig:M-R hhe}}
\end{figure}

With the same method, we derive constraints on the amounts of hydrogen
and helium that may be present. Figure~\ref{fig:M-R hhe} shows that
the presence of an envelope hydrogen and helium leads to a
very significant increase in the size of the planet.
Because of the low gravity and high
compressibility of the envelope, we find that planets with smaller
masses have larger radii if they contain a H-He envelope and are
significantly irradiated, \modif{except when the envelope to core
  mass ratio is so small that the envelope is still tightly
  bound by gravity to the Earth-like nucleus. We derive that any hydrogen-helium
  envelope in CoRoT-7b must be less than $0.01\%$ of the
  total planetary mass. Note that for such small envelopes, the
  structure is isothermal (the envelope to silicate transition occurs
  at a temperature that is within a few Kelvins of the assumed 10 bar
  temperature). If CoRoT-7b would now possess such an envelope,
  it would evaporate in only $1\,$Ma. We therefore estimate that 
  CoRoT-7b cannot possess a hydrogen-helium atmosphere.}

\subsubsection{\modif{Possibility of an undifferentiated structure}}

\modif{We have thus far assumed a differentiated structure (i.e., iron/rocks/volatiles or
hydrogen-helium). While it is indeed verified for relatively small and
cool planets (from Ganymede to the Earth), the question arises for
planets that may be large and hot enough for their interior to be
predominantly molten. 
}
\modif{
Molten silicate and water are known to be miscible with each other at
pressures above a few GPa and temperatures $>\sim1000$\,K
\citep{SK97,Mibe+07}.  Above $\sim$10000\,K, silicate and iron are also no longer
immiscible \citep{Stevenson:2008}.  Hydrogen, helium and water will also
mix at high enough temperatures.
Indeed, Uranus and Neptune appear to be only partially differentiated,
with an outer hydrogen-helium gaseous envelope that contains a large
abundance of at least one of the volatiles components, i.e. methane,
an inner dense envelope which appears to be mostly made of
high-pressure and high-temperature ``ices'', probably mixed with
rocks, and a central dense nucleus (probably made of rocks and/or
iron) with a mass of order $1\,\mearth$ or smaller
\citep{Hubbard:Neptune}. Some interior models compatible with
the observed $J_2$ and $J_4$ values  even include an inner envelope
composed of a mixture of hydrogen-helium and ``ices'' \citep{Marley+95}.
}

\modif{ Specific studies are required to solve this problem. We stress
  that our limits on the presence of a gaseous or vapor envelope
  cannot be applied as constraints on the presence of these gases
  and/or volatiles in the mantle. If the planet is undifferentiated, a
  larger component of gases and volatiles may be present than derived
  in the previous sections. (This is because the higher mean molecular
  weight will at some point prevent the planet from inflating
  thermally as much as if all light species were present in an outer
  envelope). For robust conclusions concerning undifferentiated
  super-Earths in general, detailed models beyond the
  scope of this paper are needed. The lack of adequate equations of
  state and relevant opacities are difficult limitations to overcome.
}

\modif{
In any case, we note that the mean temperature in the atmosphere
should be low enough to maintain a low abundance of silicate species
there. Specifically, if we assume a photospheric level close to the
mean equilibrium temperature, $T_0=1800$\,K, the saturation pressure
of MgSiO$_3$, representative of rock species, for that temperature is
$P_{\rm s}\approx \exp(-58663/T+25.37)\approx 7\times 10^{-4}$\,bar
(Lunine et al. 1987), i.e. much smaller than the photospheric pressure
$P_0=(2/3)(g/\kappa)\approx 1$\,bar for
$\kappa=10^{-2}\rm\,cm^2\,g^{-1}$. We thus envision that the outer
atmosphere is rich in volatiles, leading to a preferential escape of
these even in the homogeneous interior case (see
fig.~\ref{fig:Fe+R+steam}).}

\subsubsection{CoRoT-7b as an evaporated ice or gas giant?\label{sec:evaporated}}

We now examine whether CoRoT-7b may have been formed by
outstripping a gas giant or an ice giant from its envelope, leaving a
planet with little or no gaseous envelope. In order to do so, we first
calculate an ensemble of evolution models with a
  constant total mass and a constant composition. This ensemble of models 
  is characterized by a \modif{central seed of Earth-like composition of $5\,\mearth$} and variable
  total masses (from $10$ to about $120\,\mearth$). The combined mass
  and thermal evolution of a planet with mass loss is then calculated
  by noting that for each planetary mass and
  central specific entropy corresponds a given planetary radius, and therefore a
  given mass loss, and that central entropy should be conserved during mass loss:
\begin{eqnarray}
{dM\over dt}&=&f(\rho_{\rm p}),\\
{dS_{\rm c}\over dt}&=&\left(\partial S_{\rm c}\over \partial t\right)_M,
\end{eqnarray}
where $f(\rho_{\rm p})$ is a function of the planet's density provided
by eq. ~\ref{eq: escape flux 2}, and the loss of central entropy
$S_{\rm c}$ is calculated from individual evolution models with fixed
mass. 

\modif{
Because evaporation is highly dependent on the planetary density, the
choice of initial conditions affects the results directly. Here, we
only highlight reasonable possibilities for the origin and fate
of the CoRoT-7b. We assume that our planets fill their Roche lobe
when formed, and allow them to contract for 10\,Ma before turning on
mass loss, using eq.~(\ref{eq: escape flux}) . This mimics the
fact that during the protoplanetary disk phase, the planet should be
protected from mass loss but should have begun its contraction and
loss of entropy. The result also depends on the planet's orbital
history, as a planet that is initially far from its star tends to
contract faster. We therefore examine two cases: a) One in which the
planet is assumed to form in situ and remain at its present orbital
distance (0.017\,AU) throughout its existence; b) One in which its
first 10\,Ma of existence are spend at larger orbital distances,
i.e. $\sim 0.08$\,AU (corresponding to an equilibrium 10\,bar
temperature of 1000\,K), before it is suddenly brought to its present
orbital distance (and an equilibrium 10\,bar temperature of 2500\,K.
}

\begin{figure}[htbp]
\begin{centering}
\includegraphics[width=0.47\textwidth]{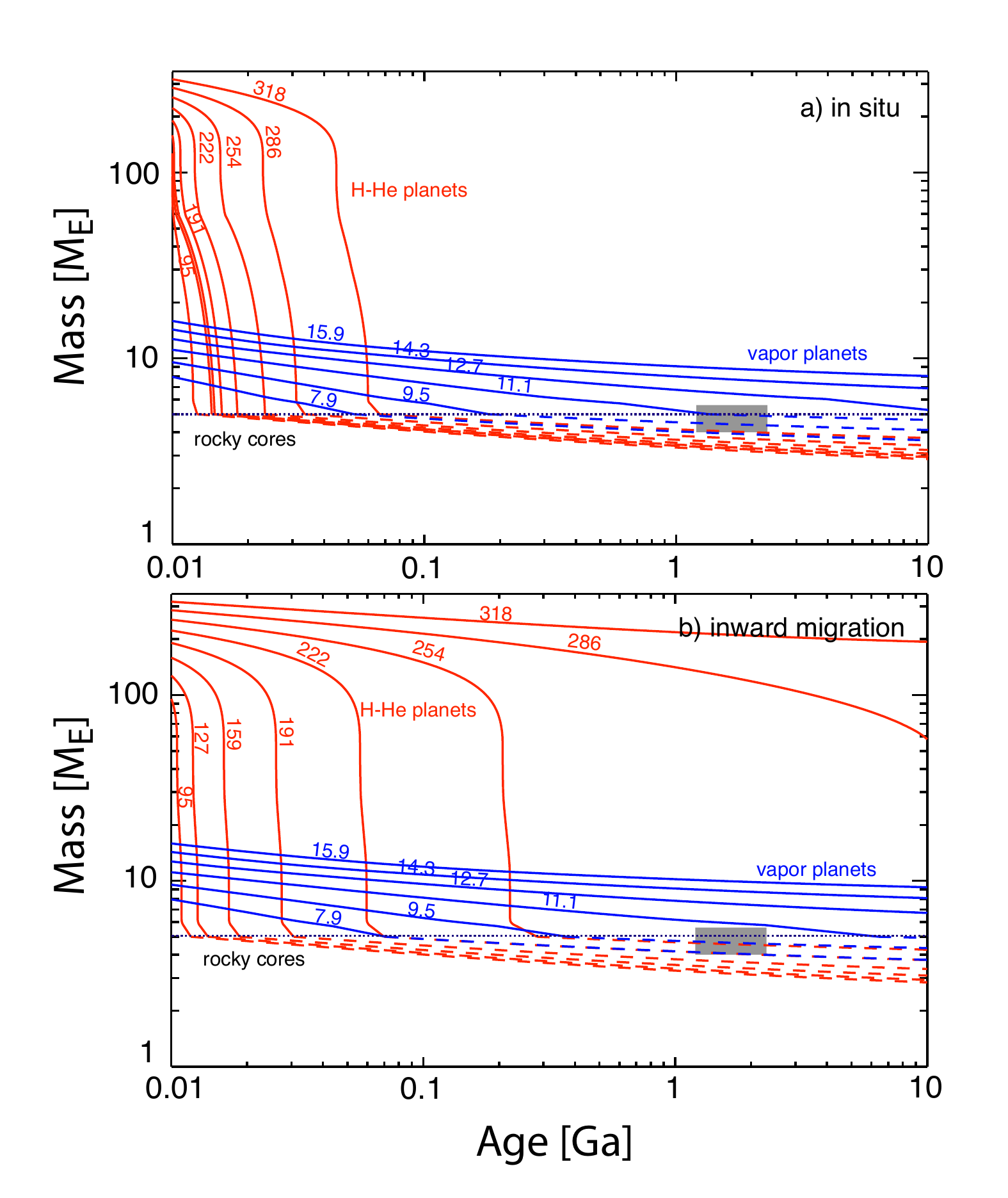}
\end{centering}
\caption{\modif{Evolution of the mass of hypothetical CoRoT-7b precursors as a
  function of time. Two type of precursors are shown: planets with an
  extended hydrogen-helium envelope (red), and planets with a vapor
  (i.e. water) envelope. Labels indicate initial masses, in Earth
  masses. All cases assume an inner $5\,\mearth$
  Earth-like core. Plain lines indicate planets that still possess an
  extended envelope. Dashed lines correspond to (evaporating) bare
  rocky cores. The top panel assumes no migration of the planet
  (maximal mass-loss). The bottom panel assumes a contraction of an
  initially exented planet during $10\,$Ma at $\sim 0.08$\,AU before
  migrating to its present location. The present mass and age of
  CoRoT-7b is indicated as a grey rectangle. A maximal, energy-limited
  evaporation is assumed (see text).} \label{fig:massloss}}
\end{figure}

\modif{
The results are shown in Fig.~\ref{fig:massloss}. Vapor planets tend
to be relatively compact and suffer significant but limited mass
loss. For example, present observations would be compatible with a
vapor planet that was initially about $12\,\mearth$ and that lost 97\%
of its vapor envelope in the ``in situ'' case. For this type of
planet, the situation remains very similar when considering the ``inward
migration'' scenario: the additional cooling only makes the vapor
envelope slightly more compact so that the ``ideal'' precursor mass
decreases to about $10\,\mearth$.
}
\modif{
The situation is different for hydrogen-helium planets because as was
shown in fig.~\ref{fig:M-R hhe} they tend to be very tenuous and
loosely bound to the rocky nucleus. For example, we find that for planets
at 0.017\,AU filling their Roche lobe, the evaporation proceeds faster
than the contraction and always lead to the complete loss of the
envelope, even for initial masses as high as the mass of Jupiter. The
top panel of fig.~\ref{fig:massloss} shows that even after 10\,Ma of
evolution without mass loss, a Jupiter-mass planet is stripped of its
entire envelope in less than 80\,Ma. If the planet is allowed to
contract at larger orbital distances (our ``inward migration'' case),
the resulting planet is more compact, and possible precursors to
CoRoT-7b are planets less massive than about $270\,\mearth$. Thus, this
shows that the evolution of layered planets maybe very different and
lead to a much more significant mass loss than when considering that
the planets contract with simple power laws \citep{Lammer09}.
Of course, the numbers that we derived may vary quite significantly depending on the
chosen formation scenario and dynamical evolution of the planet, but
at least they show that CoRoT-7b could have lost tens of Earth masses
in hydrogen and helium, and that there is no impossibility that it was
initially a gas giant.
}

\modif{
Our models are based on the assumption of the presence of distinct
layers. An undifferentiated planet would evaporate at quantitatively
different rate but we believe that qualitatively the situation would
be very similar: it would lose preferentially its light elements
(gases and volatiles), and at a rate that would be directly related to
the proportions of hydrogen and helium and volatiles that the planet
contains. 
}

From the point of view of interior and evolution models, there is thus
a range of possibilities to explain the characteristics 
of CoRoT-7b: it may have been initially a gas giant planet
that was eroded down to its dense rocky interior, it may have been a Uranus-like ``ice
giant'' which would have lost most or all of its volatiles, or it may have been all the way a rocky planet with no ice or hydrogen and
helium.

\section{Conclusion}

CoRoT-7b is the first of possibly many extreme close-in super-Earths that will be discovered in the near future.  
We have shown that the atmospheric escape for this type of planets is expected to be high, within an order of magnitude of that of HD~209458~b, and mostly independent of composition. A simple analysis shows that for CoRoT-7b, the mass already lost to escape would be of order \modif{$\sim 4 \rm M_{\oplus}$} for silicate-iron planets, or $\sim 10-100 \rm M_{\oplus}$ if the planet initially contained a massive water (hydrogen-helium) envelope.

\modif{
Given the observational constraints on its size and mass,
  CoRoT-7b is best fitted by a rocky planet that would be
  significantly depleted in iron relative to the Earth. Such a massive
  ``super-Moon'' is very unlikely to form. However, a one sigma
  increase in mass ($5.6\,\mearth$) and one sigma decrease in size
  ($1.59\,\rm R_\oplus$) would make the planet compatible with an
  Earth-like composition (33\% iron, 67\% silicates). Such a rocky planet
  would have a thin vapor silicate atmosphere. We estimate that this
  atmosphere should be thick enough to capture efficiently stellar UV
  photons therefore yielding a significant mass loss, possibly close
  to the rate obtained from energy-limiting considerations.
}
\modif{
Another possibility is that the planet was initially an ice
giant that lost most of its volatile content. We calculate the maximum
volatile envelope (in the form mostly of water vapor) for
  CoRoT-7b to be $10\%$ by mass, with a best fit solution at $3\%$. We
  also constrain a possible hydrogen-helium outer envelope to be less
  than 0.01\% by mass. In both cases, these numbers are derived
  assuming the presence of an inner rock/iron core of Earth-like
  composition. The precursor of such a planet may be a Uranus-like ice
  giant, or it may have been a more massive gaseous planet: this is
  because gaseous planets at very short orbital distances may be
  extremely extended and have their envelope only weakly bound to the
  central rock core. 
}
\modif{
Given the fast planetary erosion, we estimate that the vapor
  envelope that we derive to fit the present models would last up to
  1\,Ga before its complete erosion. In the case of a hydrogen-helium
  envelope, its survival time is only 1\,Ma. This therefore implies
  that CoRoT-7b cannot contain a hydrogen-helium envelope. It is the
  first time that such a conclusion can be drawn for an exoplanet.
}
\modif{A caveat is, however, that these estimates are based on the hypothesis that the
  structure of the planet is in the form of distinct layers with the
  light species on top. It may not be the case, in particular because
  water and silicates appear to mix efficiently at modest temperatures
  and pressures (above 1000K and a few GPa), in which case potentially
  more volatiles may be ``hidden'' in the planetary interior. This is
  probably not a concern for hydrogen and helium because they should
  be mostly on top of the rocky/icy nucleous at the time of the planet
  formation already. With specific models, we have shown that the
  undifferentiation of iron with respect to silicate does not affect
  significantly the planetary structure either. When it comes to water
  and silicates, the effect may be more pronounced because of
  their different densities and thermal properties. Specific studies
  including proper equations of state and opacities are required to
  progress in that respect.}

In parallel, the mass loss that we infer implies that there is a possibility to probe for the composition of the outer shell of the planet by measuring the composition of the extended planetary exosphere. CoRoT-7 ($V=11.7$) is significantly fainter than HD209458 ($V=7.65$), so that the measurement is challenging. However, the fact that a detection of escaping H, C and O is possible for HD~209458~b \citep{VM03, VM04} yield great hopes for similar measurements for CoRoT-7b and close-in super-Earths in the nearby future.

\begin{acknowledgements}

  This research was carried out as part of a Henri Poincare Fellow at
  the Observatoire de la C\^ote d'Azur to DV. The Henri Poincare
  Fellowship is funded by the CNRS-INSU, the Conseil General des
  Alpes-Maritimes and the Rotary International -District 1730.  MI got
  financial support from Program for Promoting Internationalization of
  University Education from the Ministry of Education, Culture,
  Sports, Science and Technology, Japan, and from the \emph{Plan
    Pluri-Formation} OPERA. The authors acknowledge CNES and the CNRS
  program \emph{Origine des Plan\`etes et de la Vie} for support. The
  authors further thank the CoRoT community, Guillaume Morard, Didier
  Saumon, Bruce Fegley and Helmut Lammer for discussions and sharing
  results in advance of publication, and the anonymous reviewer for
  helping to improve this manuscript.
\end{acknowledgements}

\end{document}